\documentclass[smallextended,numbook,envcountsame,envcountreset,natbib]{svjour3}

\usepackage{graphicx}
\usepackage{dcolumn}
\usepackage{bm}
\def\hatt{}   
\usepackage{chngcntr}
\counterwithout{equation}{section}
\def\OO{\mathcal{O}}
\let\epsilon\varepsilon
\newcommand{\ie}{{\it i.e.,\ }}

\newcommand{\be}{{\bf e}}
\newcommand{\bt}{{\bf t}}
\newcommand{\bn}{{\bf n}}
\newcommand{\bx}{{\bf x}}
\newcommand{\by}{{\bf y}}
\newcommand{\bz}{{\bf z}}

\newcommand{\br}{{\bf r}}

\newcommand{\undephi}{\rm \underline{e^{i\phi}}}
\newcommand{\undi}{{\rm \underline{i}}}
\newcommand{\undzr}{{\rm \underline{0}}}
\newcommand{\undon}{{\rm \underline{1}}}

\newcommand{\D}{{D}}
\let\phi=\varphi
\def\fref#1{Fig.~\ref{#1}}
\def\C#1{\begin{center}#1\end{center}}
\def\eref#1{(\ref{#1})}
\def\Label#1{}
\usepackage{mhequ}
\usepackage{sub_JP}
\usepackage{times}
\usepackage[usenames,dvipsnames]{color}

\def\d{{\rm d}}

\journalname{}

\begin{document}
\title{Imaging in Reflecting Spheres}
\author{Jean-Pierre Eckmann, Gemunu H. Gunaratne, Jason Shulman \& Lowell T. Wood}
\institute{J.-P. Eckmann\at D\'epartement de Physique Th\'eorique and
Section de Math\'ematiques\\ Universit\'e de
Gen\`eve, 1211 Geneva 4, Switzerland \and Gemunu H. Gunaratne, \at Department of Physics, University of Houston, Houston, TX 77204 \email{gemunu@uh.edu} \and
Lowell T. Wood\at Department of Physics, University of Houston, Houston, TX 77204
\and Jason Shulman\at Department of Physics, University of Houston, Houston, TX 77204 and Department of Physics, Stockton University, Galloway, NJ, 08205}
\date{\today}
\maketitle           

\begin{abstract}
We study the formation of images in a reflective sphere in three configurations using caustics of the field of light rays. The optical wavefront emerging from a source point reaching a subject following passage through the optical system is, in general, a Gaussian surface with partial focus along the two principal directions of the Gaussian surface; \ie there are two images of the source point, each with partial focus. As the source point moves, the images move on two surfaces, referred to as \emph{viewable surfaces}.  In our systems, one viewable surface consists of points with radial focus and the other consists of points with azimuthal focus. The problems we study are (1) imaging of a parallel beam of light, (2) imaging of the infinite viewed from a location outside the sphere, and (3) imaging of a planar object viewed through the point of its intersection with the radial line normal to the plane. We verify the existence of two images experimentally and show that the distance between them agrees with the computations.   
\end{abstract}


\section {Introduction}

\begin{figure}
\C{\includegraphics[width = 0.60\textwidth]{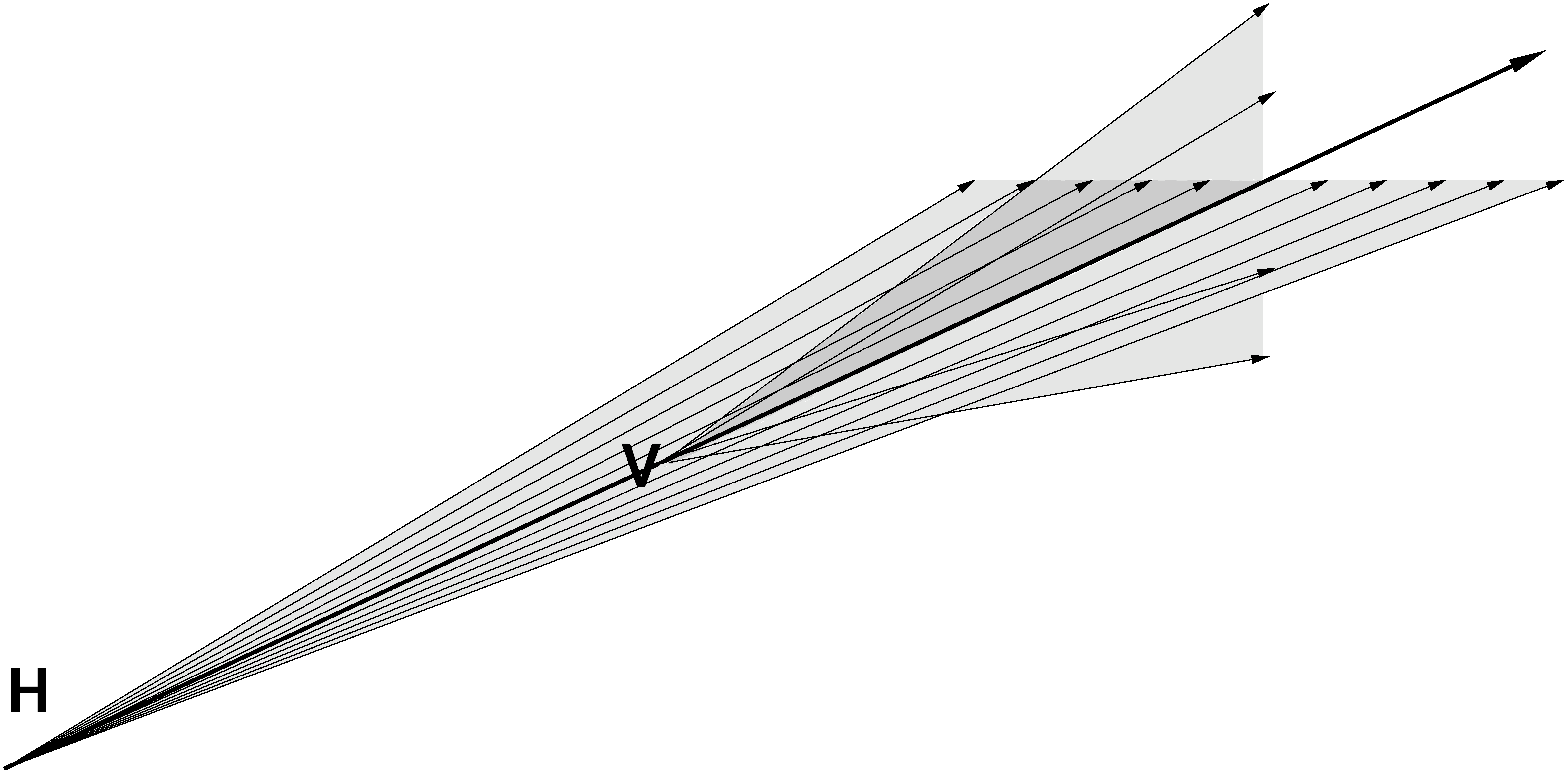}}
\caption {An optical wavefront is a Gaussian surface defined by two principal directions and the associated radii. The only partial focus of rays is along these principal directions. The two families of rays appears to emerge from ``images" at $H$ and $V$. Reproduced from \citep{Eckmann:2021} .}
\label{fig:Fan}
\end{figure}

In Chapter 3 of an as yet unpublished treatise on visual perception~\citep{Feigenbaum2050}, the late chaos-theorist M.J.~Feigenbaum 
discussed formation of images in optical systems, which form at caustics of the field of light rays~\citep{Nye:1999,Berry2015,Nussenzveig1977}. Specifically, a small section of a spherical wavefront emerging
from an object will evolve during its passage through an optical system  
to a Gaussian surface characterized by two orthogonal principal directions and the associated radii. Partial focusing of rays
is only achieved along these principal directions as illustrated in \fref{fig:Fan}~\citep{Eckmann:2021,Kinsler:1945,BartlettLucero:1984,HorvathBucha:2003}. The wavefront can be represented using the two families of rays
(here chosen to be vertical and horizontal)  appearing to emerge
from a pair of distinct ``non-objects" $V$ and $H$. On focussing at $V$, the horizontal direction will appear unfocussed, the view being referred to as \emph{horizontal astigmatism}. Similarly, focussing on $H$, the vertical direction will appear unfocussed, resulting in \emph{vertical astigmatism}.   Although the visual system is capable of focusing on either image,
Feigenbaum argued that it naturally selects the vertically astigmatic image.  However, unlike prior conjectures~\citep{HorvathBucha:2003}, he showed that this choice is not predicated on binocularity, by requesting subjects to report what is seen with one eye closed. 

Ref.~\citep{Feigenbaum2050} illustrates  the formation of anamorphic images, \ie images formed in a cylindrical reflector of a horizontal table. One of the two images of the table was shown to lie inside the cylinder while the other lay on the table behind the cylinder. The two surfaces on which the table was imaged were referred to as \emph{viewable surfaces}. The viewable surface inside the cylinder consisted of points with horizontal focus (such as $H$) while that on the table behind the mirror contained points with vertical focus. As predicted, it was found that in erect viewing, subjects observe the image within the mirror while in orthogonal viewing (with eyes parallel to the mirror axis), they observe the image on the table.

In this paper, we study imaging from  a witch ball, \ie a reflective sphere.
Viewing through the sphere, one observes a deformed image of its surroundings. This phenomenon is well-known
and, for example, discussed in \citep{Berry:1972,Berry2015} where one of the images is 
calculated. Here, we extend the study using the approach of
  \citep{Feigenbaum2050}. These findings involve calculation of caustics~\citep{AvendanoAlejo:2013}, which are simple geometry, albeit somewhat tedious.

The calculation outlined in Section~\ref{sec:ParallelBeam} is that of locating images of a parallel beam of light
impinging on the sphere. One image surface consists of points of radial focus. It is an azimuthally symmetric surface whose cross section contains a cusp singularity. The calculation is that presented in \citep{Berry2015}. The second surface consists of points of azimuthal focus. Symmetry considerations dictate that the surface is degenerate, collapsing to the mirror axis parallel to the incoming beam.

In Section~\ref{sec:Infinite}, we consider a configuration where both viewable surfaces are non-degenerate. It is the viewpoint of a cyclopic observer at a finite distance viewing the infinite through the sphere.  
Two distinct surfaces are found \emph{inside} the sphere, distinguished by different types of astigmatism, 
radial and azimuthal. It is found that if the viewing distance is approximately 3.73 times the radius of the sphere, the radially focussed viewing surface is planar to fourth order in perturbation. It is an ideal distance to capture a photographic image of the infinite viewed in the sphere. 
  
 A second non-degenerate problem considered in Section~\ref{sec:ViewingPlane} consists of viewing a plane through the point of intersection of a mirror axis with the plane. Once again we find two image surfaces associated with radial and azimuthal astigmatism. In Section~\ref{sec:Expt} we demonstrate the presence of two distinct images and compare experimentally measured inter-image distances with  results from the computation.  
 
 Section~\ref{sec:Discussion} discusses implications of our results.

%
%
%
%

\vskip 0.3in

\section {Images of a Parallel Incoming Beam}
\label{sec:ParallelBeam}

\begin{figure}
\C{\includegraphics[width = 0.8\textwidth]{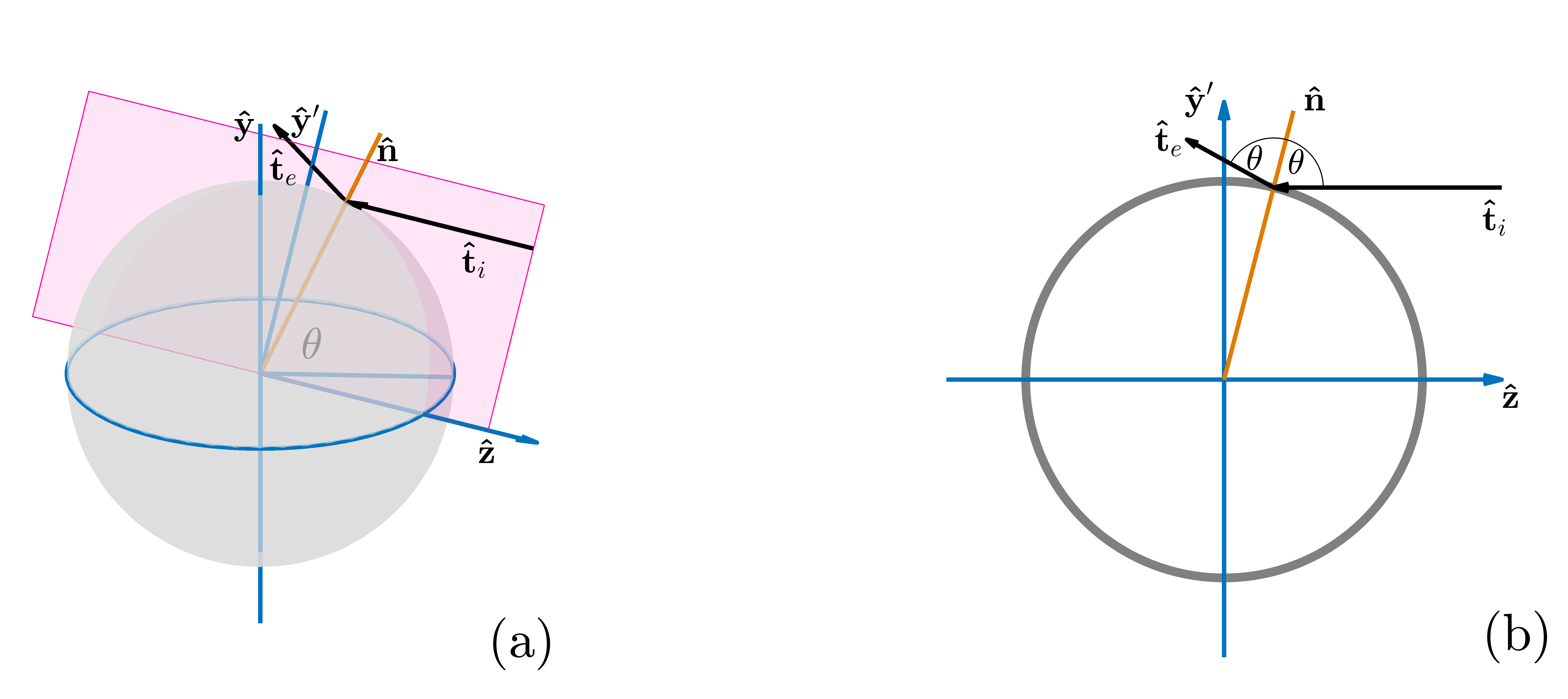}}
\caption {(a) A schematic of a ray of light (black), incident along the $\bt_i = -{\hat  \bz}$  direction, impinging on a spherical witch ball of radius $R=1$. The ray is incident on and reflects from the surface at the point defined by polar angles $(\theta, \phi)$. (b) The ray-diagram in the plane $(\bt_i,\  {\hat  \bn})$.}
\label{fig:ChristmasBall}
\end{figure}

Studies outlined in this section refer to the configuration shown
schematically in \fref{fig:ChristmasBall}(a). A beam of light,
incident in the $-{\hat  \bz}$-direction, impinges on and reflects off the sphere. 
A representative ray reflects from a point on
the witch ball defined by polar angles $(\theta, \phi)$, $\theta$ being the angle of incidence. The direction
of the incident ray $\hat \bt_i$ and the normal ${\hat  \bn}$ to the sphere at incidence are  
\begin{equ}
\hat \bt_i = [{\rm \underline{0}},\ -1], \qquad
\hat  \bn = [\sin \theta\ \undephi,\ \cos \theta]~,
\Label{eqn:incident}
\end{equ}
where the underbar in the complex first component indicates that the ${\hat  \bx}$ and ${\hat  \by}$ components are the real and imaginary parts of the complex number. In this representation the scalar product of two vectors $\left[ u_1{\rm \underline{e^{i\phi_1}}}, z_1 \right]$ and $\left[ u_2{\rm \underline{e^{i\phi_2}}}, z_2 \right]$ is $u_1 u_2 \cos (\phi_1 - \phi_2) + z_1 z_2$.

Define a second basis $\left\{ {\hat  \bx}^{\prime}, {\hat
  \by}^{\prime}, {\hat  \bz}^{\prime} \right\}$ such that ${\hat
  \bz}^{\prime} \equiv {\hat  \bz}$ and the plane of reflection is the
${\hat \bx}^{\prime}{\hat  \bz}^{\prime}$ plane as shown in  \fref{fig:ChristmasBall}. The new
basis vectors are  
\begin{equ}
{\hat  \bx}^{\prime} = \left[{\undephi}, 0 \right]~;
\ \ \ \ \ {\hat  \by}^{\prime} = \left[ {\rm \underline {i e^{i\phi}}}, 0 \right]~;
\ \ \ \ \ {\hat  \bz}^{\prime} = \left[{\rm \underline{0}}, 1\right]~.
\label{eqn:Transformation}
\end{equ}
Reflection of the ray in the ${\hat  \bx}^{\prime}{\hat  \bz}^{\prime}$ plane is shown in \fref{fig:ChristmasBall}(b). The 
outgoing ray begins at the point of incidence $\be_0 =\hat  \bn$ and is directed along
\begin{equ}
\hat \bt_e = \sin 2\theta \ {\hat  \bx}^{\prime} + \cos 2\theta\ {\hat  \bz}^{\prime} 
         = \left[ \sin 2\theta\ {\undephi}, \ \cos 2\theta \right].
\end{equ}
Hence, the outgoing ray can be parameterized as
\begin{equ}
\br(s; \theta, \phi) =  \be_0 + s \cdot \hat \bt_e = \left[ (\sin \theta + s \sin2\theta)\ {\undephi},\ \cos \theta + s \cos 2\theta \right]. 
\label{eqn:reflectedray}
\end{equ}

In \citep{Feigenbaum2050}, it is highlighted that images are formed on caustic surfaces of a wavefront, and can be computed using a collection of neighboring exiting rays. The caustics, envelopes of neighboring rays, are given by $\d \br (s) = 0$; \ie
\begin{equ}
\frac{\partial}{\partial \theta} \br(s; \theta, \phi) \d \theta + \frac{\partial}{\partial \phi} \br(s; \theta, \phi) \d \phi + \frac{\partial}{\partial s} \br(s; \theta, \phi) \d s = 0~.
\label{eqn:derivatives1}
\end{equ} 
Since the configuration is axially symmetric, we can select a cone of rays around $\phi=\pi/2$ (\ie on the vertical axis) to evaluate the images. Using \eref{eqn:reflectedray},
\begin{equa}
\d \br =& \left[ (\cos \theta + 2 s \cos 2\theta)\ \undi, \ -\sin \theta - 2 s \sin2\theta \right] \d\theta \\
   &+ \left[-(\sin \theta + s \sin 2\theta)\ \undon,\ 0 \right] d\phi + \left[\sin 2\theta \ \undi,\ \cos 2\theta \right] \d s = 0~,
\label{eqn:reflectedray2}
\end{equa}
whose ${\hat  \by}$ and ${\hat  \bz}$ components are
\begin{equa}
(\cos \theta + 2 s \cos 2\theta)\ \d\theta + \sin 2\theta \ \d s &= 0~, \\
-(\sin \theta + 2 s \sin 2\theta)\ \d\theta + \cos 2\theta \ \d s &= 0~. \\
\end{equa}
Setting the determinant of coefficients to zero gives 
\begin{equ}
\left( \cos \theta + 2s \right) \ \d\theta = 0~.
\label{eqn:condition1}
\end{equ}
The ${\hat  \bx}$ component of \eref{eqn:reflectedray2} is
\begin{equ}
\left(\sin \theta + s \sin 2\theta \right)\ \d\phi = 0~.
\label{eqn:condition2}
\end{equ}
Equations \eref{eqn:condition1} and \eref{eqn:condition2} can be used to calculate locations of the images at $\phi=\pi/2$.

\begin{figure}
\C{\includegraphics[width = 0.60\textwidth]{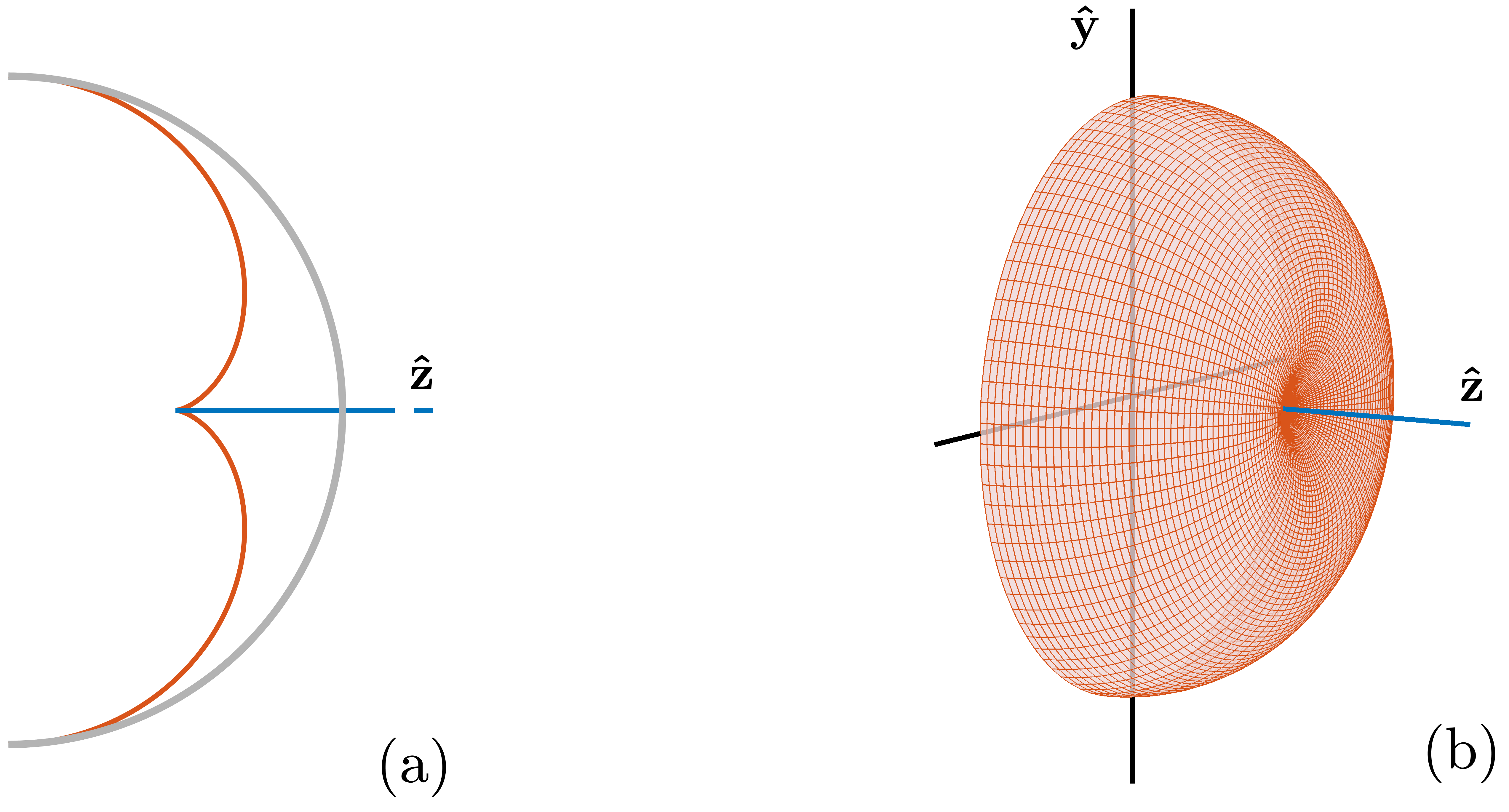}}
\caption {The two caustics for reflection of a beam of light parallel
  to the $z$-axis from a spherical mirror. (a) The $R$-caustic, consisting of 
  points from which radial fans of rays (appear to) emerge, is the
  cross section of the surface shown in red in (b). The cross
  section of the mirror is shown in gray. The blue line shows the degenerate
  $A$-caustic $[\frac{1}{2}, \infty)$ on the $z$-axis.}
\label{fig:CausticsVH}
\end{figure}

One solution to equations \eref{eqn:condition1} and \eref{eqn:condition2} satisfies $\d\phi=0$ implying $\d \theta \ne 0$ (for otherwise we only have a single ray), and $s = s_R = -\frac{1}{2} \cos \theta$. The location of this image is 
\begin{equ}
\br_R = \left[\left(\frac{3}{2}\cos \theta - \cos^3 \theta \right)\ \undi, \ \sin^3 \theta \right].
\label{eqn:RSoln}
\end{equ}
The resulting surface, which we refer to as the $R$-caustic, consists of points of radial focus and azimuthal astigmatism. It is shown in red in \fref{fig:CausticsVH} and contains a singularity on the $z$-axis. To establish the nature of the singularity, note that $y_R - 1/2 \sim \theta^2$ and
$z_R \sim \theta^3$, and hence the singularity is of power
$3/2$; \ie a cusp singularity. This caustic is discussed in \citep{Berry2015}. 

The second solution is given by $\d\theta = 0$ and $\d\phi \ne 0$. From \eref{eqn:condition2}, the value of $s$ for this image is $s_A = -1 / 2 \cos \theta$, and the location of the image (from \eref{eqn:reflectedray}) is
\begin{equ}
\br_A = \left[\undzr, \frac{1}{2\cos \theta} \right] = \frac{1}{2 \cos \theta}\  {\hat  \bz}.
\label{eqn:ASoln}
\end{equ}
The surface containing points of azimuthal focus (and radial astigmatism) is called the $A$-caustic; it lies on the $\bz$ axis at distances $[\frac{1}{2},
\infty)$ and is shown in blue in \fref{fig:CausticsVH}.

The degeneracy of the $A$-caustic originates from the azimuthal symmetry of the configuration. This is most easily deduced by noting that the exit ray reflected from a point defined by angles $(\theta, \phi)$ lies in the plane defined by $\phi$. Hence two rays impinging at $\phi$ and $\phi+d\phi$ will meet on the intersection of the associated planes, which is the $z$-axis.  

Note that one caustic was given by $\d \phi=0$. This was required by reflection symmetry of the configuration about the $y$-axis. Since the two caustics are orthogonal, the second is required to satisfy $\d \theta=0$. We will use these symmetry requirements to simplify computations in the next two sections

\section {Viewing the Infinite through the Witch Ball}
\label{sec:Infinite}

Consider cyclopic viewing of the witch ball from a point $O$ at a
distance $D$ from its center, assumed without loss of generality to be
on the $z$-axis. Points imaged are infinitely far away from the witch
ball. Consider a ray of light arriving in a direction $\hat \bt_i$ at the
point ${\hat  \bn}=\left[ \sin \theta\ \undephi,\ \cos \theta \right]$, on the mirror, and reflecting in a direction
$\hat \bt_e$ to reach the point $O= \left[ \undzr, \D \right]$. We search for images seen from $O$ 
of all infinitely far away points. 

\begin{figure}
\C{\includegraphics[width = 0.50\textwidth]{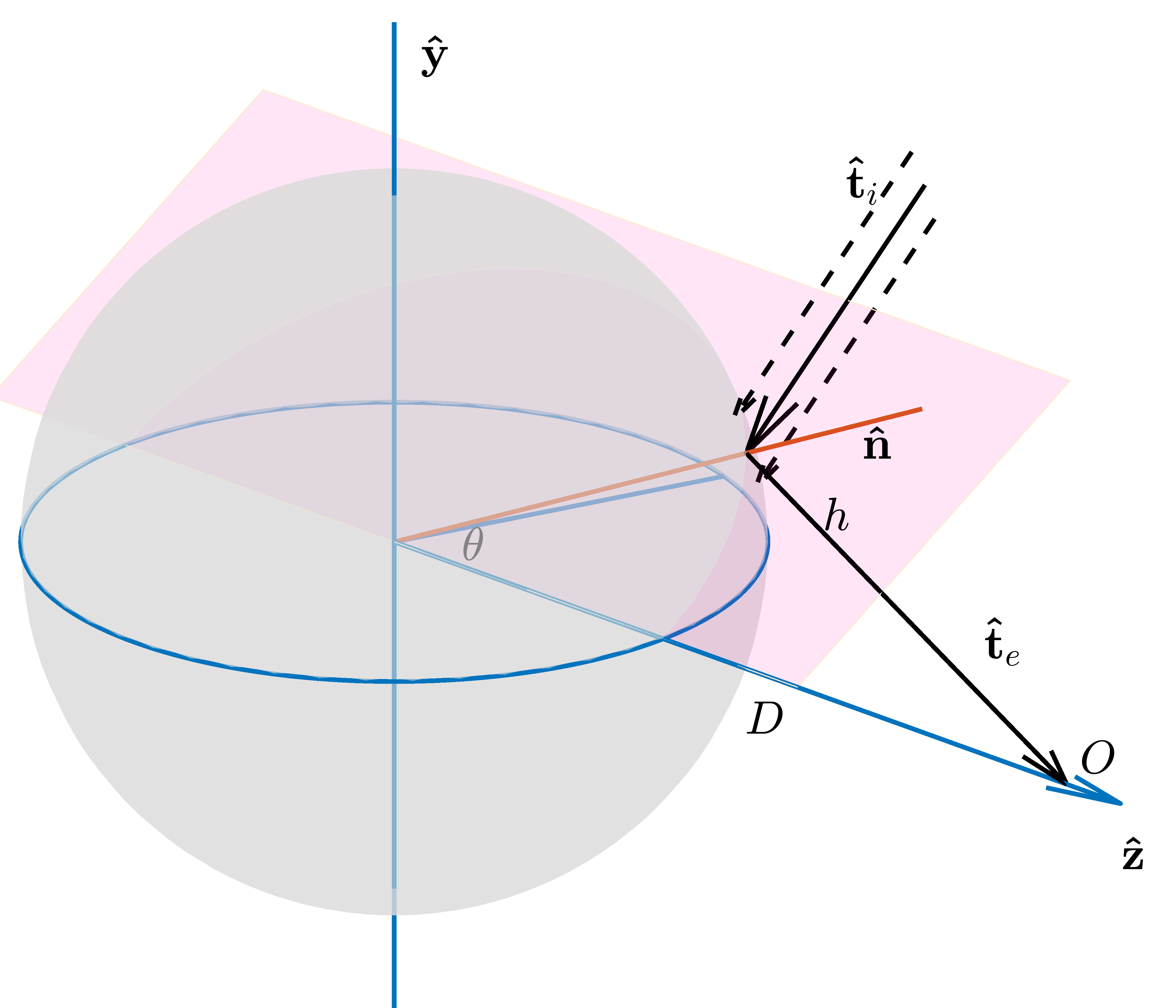}}
\caption{The ray diagram for beam arriving in a direction
  $\hat \bt_i$ to the point ${\hat  \bn}$, on the mirror, and reflecting in
  a direction $\bt_e$ to reach $O= \left[ \undzr, \D \right]$. It is also necessary to
  follow reflections of a collection of rays parallel to $\hat \bt_i$ that
  are incident on points near ${\hat  \bn}$. 
    }
\label{fig:CausticsVH2}
\end{figure}

Note first that, by the laws of reflection and time-reversal symmetry (\ie
$\hat \bt_i \rightarrow -\hat \bt_e$; $\hat \bt_e \rightarrow -\hat \bt_i$), the unit vectors $\hat \bt_i$, $\hat \bt_e$, and ${\hat  \bn}$ along a ray of incidence, ray of reflection, and the surface normal satisfy
\begin{equ}
\hat \bt_e = -2({\hat  \bn} \cdot \hat \bt_i) {\hat  \bn} + \hat \bt_i;\ \ \ \ \ \ \ \ \ \ \hat \bt_i = -2({\hat \bn} \cdot \hat \bt_e) {\hat  \bn} + \hat \bt_e~.
\label{eqn:LawofReflection}
\end{equ}
For the configuration shown in \fref{fig:CausticsVH2},
\begin{equa}
{\hat  \bn} &= \left[ \sin \theta\ \undephi, \ \cos \theta \right]~, \\
{\hat \bt}_e &= \frac{1}{h} \left[- \sin \theta\ \undephi,  D -\cos \theta \right]\\
h &= \sqrt{\D^2 + 1 - 2\D \cos \theta}~.
\Label{eqn:prelim}
\end{equa}
It follows from \eref{eqn:LawofReflection} that
\begin{equ}
\hat \bt_i = \frac{1}{h} \left[ (\sin \theta - \D \sin 2\theta)\ \undephi, \ (\cos \theta - \D \cos 2\theta) \right].
\Label{eqn:tf}
\end{equ}
(It can be verified that $\hat \bt_e = -2({\hat  \bn} \cdot \hat \bt_i) {\hat  \bn} + \hat \bt_i$.)

What is to be performed is a calculation of the image formed by a beam of rays parallel to $\hat \bt_i$ 
that impinges on the sphere in the neighborhood of $\hat \bn$. Toward this end, consider such a ray
incident on the mirror at a point defined by angles $(\theta+\mu)$ and $\phi+\nu)$, for small $\mu$
and $\nu$. The normal at its point of incidence is
\begin{equ}
{\hat  \bn}^{\prime} = \left[ \sin (\theta+\mu)\ {\rm \underline{e^{i(\phi+\nu)}}},\ \cos(\theta+\mu) \right],
\Label{eqn:newNormal}
\end{equ}
and the ray reflects in a direction $\hat \bt_e^{\prime} = -2 ({\hat 
  \bn}^{\prime} \cdot \hat \bt_i) {\hat  \bn}^{\prime} + \hat \bt_i$.

  As the incident direction changes, the contact with the caustic
  slides along the caustic, with potential jumps from one smooth section to
  another one, for example from the top red arc to the bottom red
  arc in \fref{fig:CausticsVH}. The surface containing all such points is
 referred to as the ``viewable surface'' in \citep{Feigenbaum2050}. Notice that viewable surfaces depend both on the 
 caustic and the objects viewed.
 Below, we compute these surfaces
  associated with the $R$- and $A$-caustics. Due to azimuthal symmetry, the images for only a single value of 
  $\phi$, selected to be $\phi=\pi/2$, needs to be calculated. 
  

\subsection {The $R$-Viewable Surface} 

\begin{figure}
\C{\includegraphics[width = 0.60\textwidth]{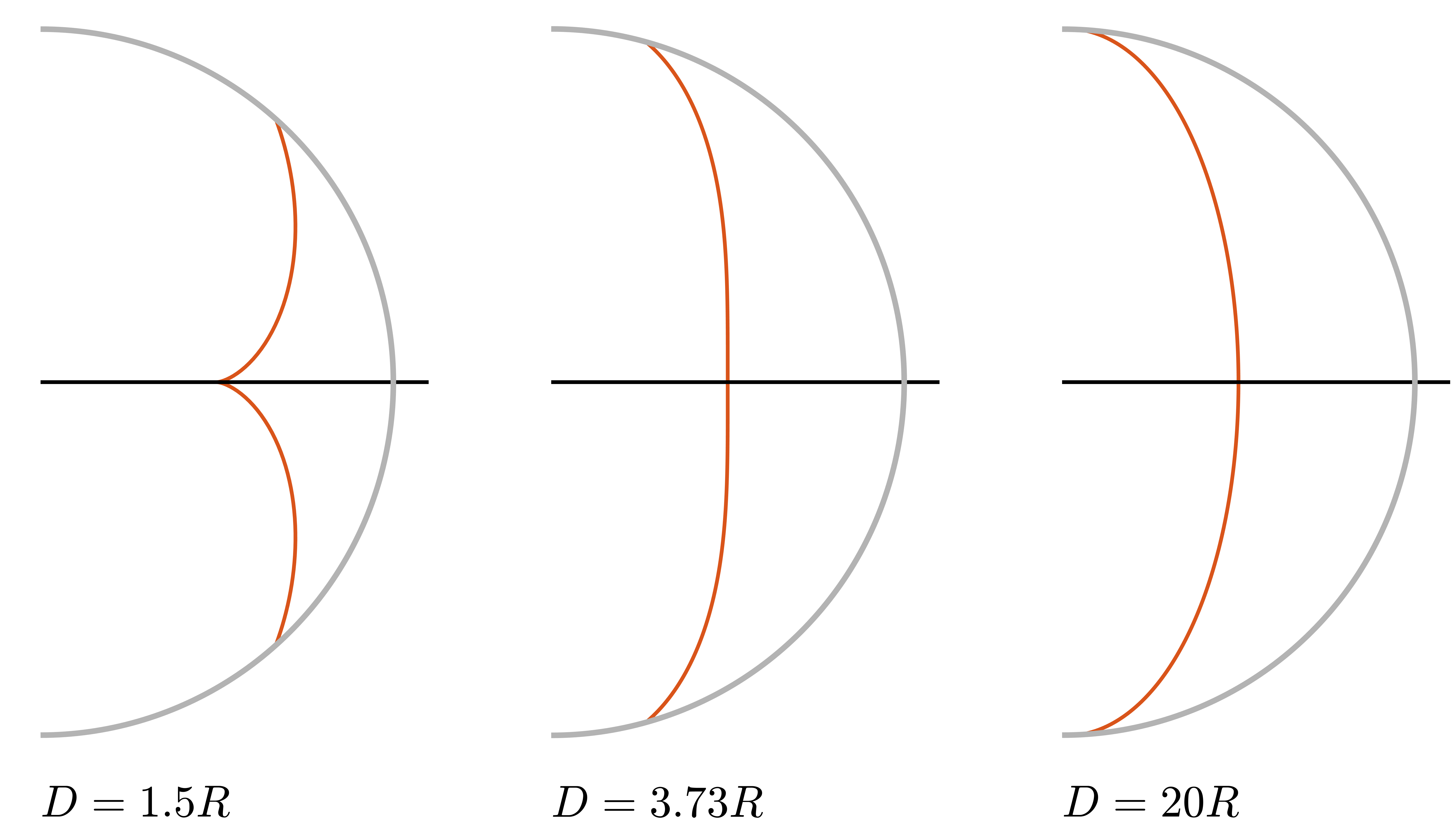}}
\caption {Cross sections of the $R$-caustics for three values of the
  viewing distance $D$.}
\label{fig:RCausticSurf2}
\end{figure}

The $R$-viewable surface consists of locations of radial focus of far-away points and satisfies $\nu = 0$. With
$\phi=\pi/2$, we find ${\hat  \bn}^{\prime} = \left[ \sin (\theta + \mu)\ \undi,\ \cos  (\theta + \mu) \right]$, and
consequently
 \begin{equ}
 {\hat  \bn}^{\prime} \cdot \hat \bt_i = \frac{1}{h}  \left\{ \cos \mu - D \cos (\theta - \mu) \right\}. 
 \end{equ}
 From \eref{eqn:LawofReflection} we get
\begin{equ}
\hat \bt_e^{\prime} = -2 ( {\hat  \bn}^{\prime} \cdot \hat \bt_i) \left[ \sin (\theta + \mu) \ \undi,\ \cos (\theta + \mu) \right] + \frac{1}{h} \left[ (\sin \theta - D \sin 2\theta)\ \undi,\ (\cos \theta -D\cos  2\theta) \right].
\end{equ}
The exit ray is $\br_R(\mu, t) = {\hat  \bn}^{\prime} + s \cdot \hat
\bt_e^{\prime}$, and the $R$-viewable surface is given by 
\begin{equ}
\d \br^{\prime} = \left( \frac{\partial \br^{\prime}}{\partial \mu} \right)_{\mu=0} \d\mu + \left( \frac{\partial \br^{\prime}}{\partial s} \right)_{\mu=0}   \d s = 0~.
\end{equ}
Noting that 
\begin{equa}
\left( \frac{\partial \hat \bn^{\prime}}{\partial \mu} \right)_{\mu=0} &= \left[ \cos \theta \ \undi, -\sin \theta \right]~,\\
 \hat \bt_e^{\prime} \vert_{\mu=0} &= \frac{1}{h} \left[-\sin \theta \ \undi, (D-\cos \theta) \right]~,\\
\left( \frac{\partial \hat \bt_e^{\prime}}{\partial \mu} \right)_{\mu=0} &= \frac{2}{h} \left[ (D - \cos \theta)\ \undi,\ \sin \theta \right]~,
\end{equa}
we find that the $x$-projection is identically satisfied, while the $y-$ and $z-$ projections give
\begin{equa}
\left\{  \cos \theta + \frac{2s}{h} (D - \cos \theta) \right\}\ \d\mu - \frac{\sin \theta}{h}\ \d s &= 0\ ,\\
  -\left(1 - \frac{2s}{h} \right) \sin \theta \ \d\mu + \frac{(D -  \cos \theta)}{h}  \ \d s &= 0. 
\end{equa}
Consistency of the pair of equations yields
\begin{equ}
s_R = -\frac{1}{2} \frac{(D \cos \theta - 1)}{h},
\end{equ}
whereupon the location of the image is found to be
\begin{equa}
\br_R &= {\hat  \bn} + s_R\cdot \hat \bt_e\label{eqn:Rcaus2} \\
       &= \left[ \sin \theta \left(1 - \frac{(D-\cos\theta)}{h^2} \right) \ \undi, \ \cos \theta - \frac{(D \cos \theta - 1) (D - \cos \theta)}{2h^2}  \right].
\end{equa}
\fref{fig:RCausticSurf2} shows cross sections of the caustic surface for
$D=1.5R$, $D=3.73R$ and $D=20R$. Note that the surface is nearly flat
for the intermediate configuration, as seen from
\fref{fig:RCausticSurf2}, an extremely good surface for sharp
imaging. 

The shape of the viewable surface near the axis of symmetry is
\begin{equ}
\left( z_R - \frac{1}{2} \right) = \frac{1}{2} \left( \frac{D}{(D-1)^2} - \frac{1}{2} \right) \theta^2 +\OO(\theta^4); \ \ \ \ \ \ \ \ y_R = \frac{(D-2)}{(D-1)} \theta + \OO(\theta^2)~,
\end{equ}
showing that the surface $z_R = z_R(y_R)$ is quadratic except when $D= 2+ \sqrt{3} \approx 3.73$, when it is quartic.

\subsection{The $A$-Viewable Surface}
The $A$-viewable surface consists of locations of azimuthal focus satisfying $\mu = 0$. With
$\phi=\pi/2$, we find ${\hat  \bn}^{\prime} = \left[ \sin \theta\ {\rm \underline{ie^{i \nu}}},\ \cos  \theta  \right]$, and
consequently
 \begin{equ}
 {\hat  \bn}^{\prime} \cdot \hat \bt_i = \frac{1}{h}  \left\{ (1-D \cos \theta) + (\cos \nu -1) \Gamma \right\}, 
 \end{equ}
 where  $\Gamma = \sin\theta (\sin \theta - D \sin 2\theta)$. From \eref{eqn:LawofReflection}
\begin{equ}
\hat \bt_e^{\prime} = -2 ( {\hat  \bn}^{\prime} \cdot \hat \bt_i) \left[ \sin \theta\ {\rm \underline{i e^{i\nu}}},\ \cos \theta \right] + \frac{1}{h} \left[ (\sin \theta - D \sin 2\theta)\ \undi,\ (\cos \theta -D\cos  2\theta) \right]~.
\end{equ}
The exit ray is $\br_A(\mu, t) = {\hat  \bn}^{\prime} + s \cdot \hat \bt_e^{\prime}$, and setting $\nu = 0$ the $A$-viewable surface is given by
\begin{equa}
0 = &\d\br^{\prime}_A = \left\{ \left[-\sin \theta\ \undon, \ 0 \right] -2 s ({\hat  \bn}^{\prime} \cdot \hat \bt_i)\left[ -\sin \theta\ \undon,\ 0 \right] \right\} \d\nu\\
                             &\left\{ -2 ({\hat  \bn}^{\prime} \cdot \hat \bt_i) \left[\sin \theta\ \undi,\ \cos \theta \right] + \frac{1}{h} \left[ (\sin \theta - D \sin 2\theta) \ \undi,\ (\cos \theta -D\cos  2\theta) \right]
 \right\} \d s ~.
 \end{equa}
 Projecting in the $y$-direction gives $\d s=0$, and projecting in the $x$-direction gives
 \begin{equ}
 s_A = -\frac{1}{2} \frac{h}{(D\cos \theta - 1)}~.
 \end{equ}
 The location of the image is
 \begin{equ}
\br_A = {\hat  \bn} + s_A \cdot \hat \bt_e = \left[ \left\{ \sin \theta + \frac{\sin \theta}{2(D\cos \theta-1)}\right\}\ \undi, \cos \theta - \frac{(D-\cos\theta)}{2(D \cos \theta - 1)} \right]~.
\end{equ}
For the $A$-viewable surface $z_A = 1/2 + \OO(y_A^2)$ at the symmetry    
  axis, where the quadratic term is a function of $D$ that does not
  vanish at any $D$.

\begin{figure}
\C{\includegraphics[width = 0.60\textwidth]{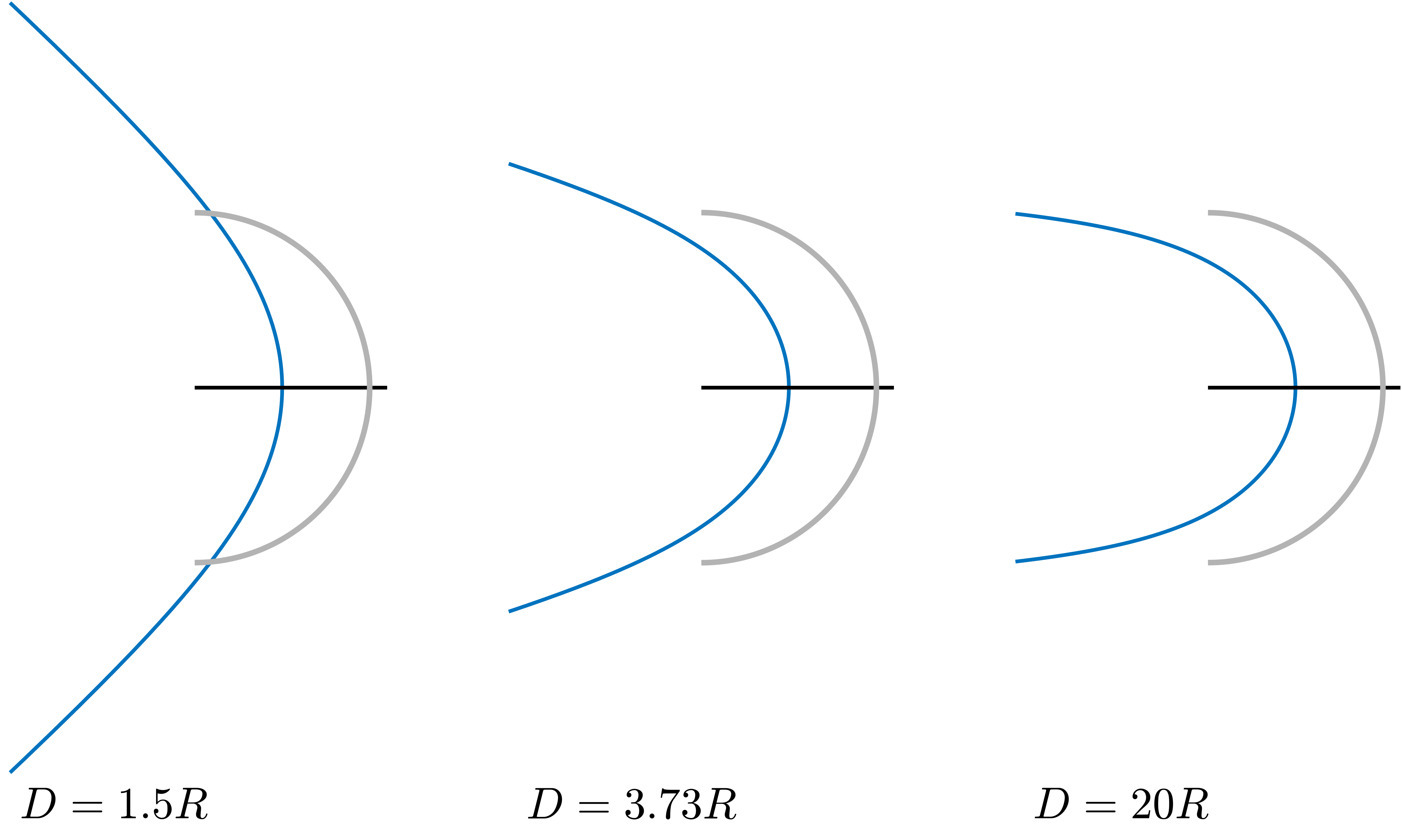}}
\caption {Cross sections of $A$-viewable surfaces for three values of the
  viewing distance $D$.
  }
\label{fig:ACaustic2} 
\end{figure}

The $z$-coordinate of the viewable surface moves to $-\infty$ as the
incoming ray reaches grazing incidence at $\theta = {\rm sin^{-1}}
(1/D)$.  \fref{fig:ACaustic2} shows cross sections of the viewable surface
for $D=1.5R$, $D=3.73R$, and $D=20R$. 

%

\section {Viewing a Planar Surface from its Center}
\label{sec:ViewingPlane}
\begin{figure}
\C{\includegraphics[width = 0.60\textwidth]{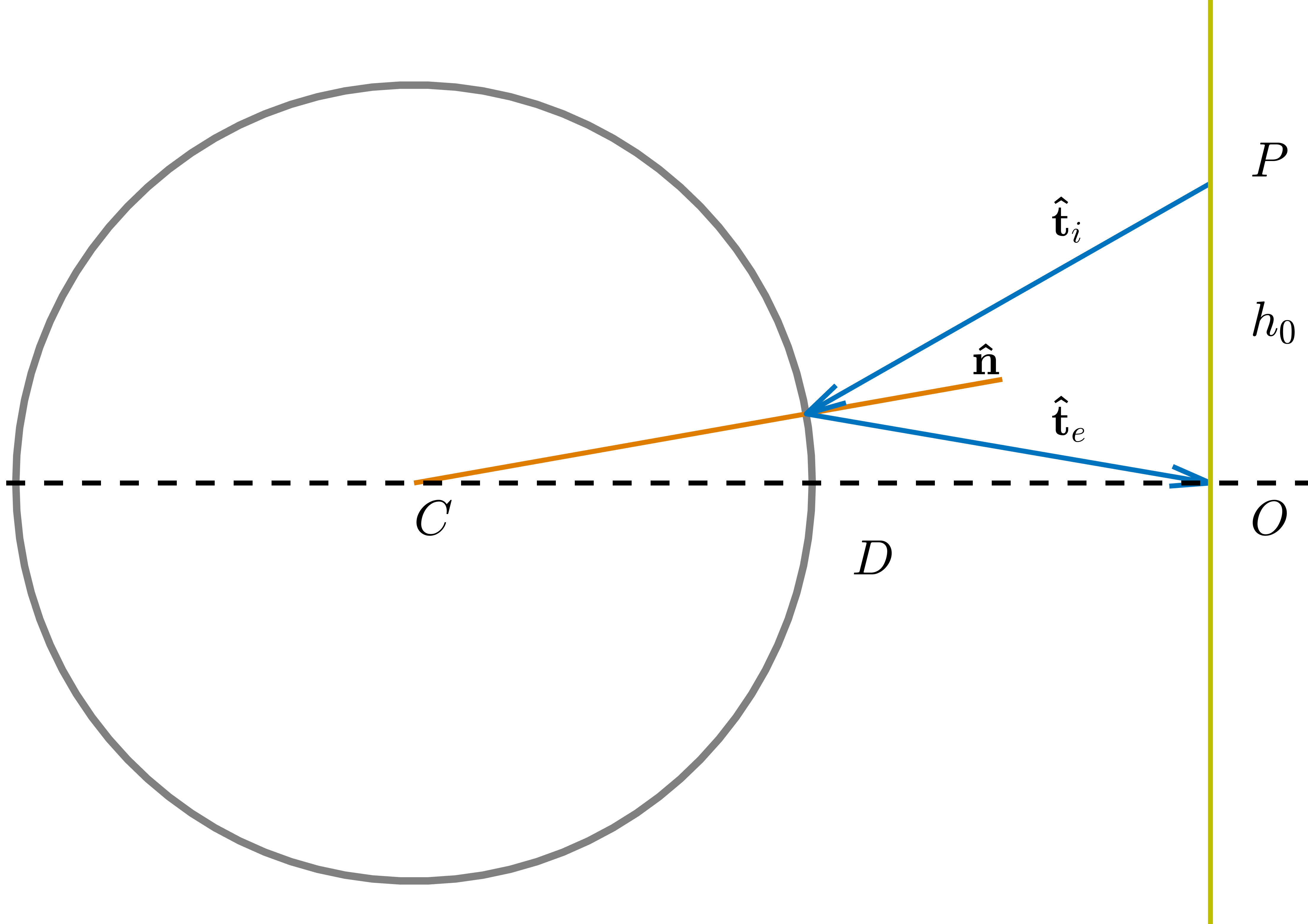} }
\caption {Viewing an image  of a plane (yellow) in the witch ball. The image is viewed from $O$, the intersection of the plane with the diameter normal to it. The ray from $P$ reaches $O$ following reflection from the ball.}
\label{fig:FlatConfig} 
\end{figure}

\fref{fig:FlatConfig} shows another configuration with non-degenerate viewable surfaces. An image,
  such as a photograph, lies on the yellow
  plane (shown in projection) a distance $D$ from the center of the
  mirror. The observer places the eye at the intersection of the black dashed
  line (looking to the left through a small hole in the center of the
  yellow surface). The reflected image can then be seen in two
  different locations, depending on the astigmatism. In \fref{fig:FlatView},
  the cross-section of the $R$-viewable surface is shown in red, and that of the $A$-viewable surface
 in blue. At the edges, focus
  will be at different depths.


The path of a representative ray, from a point $P=\left[ y_0 \undi, \ D \right]$ on the photograph to the observation point $O$ is shown in \fref{fig:FlatConfig}. Note that, since the configuration is axially
symmetric, we have, without loss of generality, chosen $\phi=\pi/2$. The direction of the incoming ray is
\begin{equ}
\hat \bt_i = \frac{1}{h_i} \left[( \sin \theta - y_0)\ \undi, \ (\cos \theta - D) \right]~,
\end{equ}
where
\begin{equa}
h_i^2 =\, & |(\sin \theta - y_0)\ \undi, (\cos \theta - D)|^2 \\
          =\, & 1 + D^2 + y_0^2 - 2D\cos \theta  - 2 y_0 \sin \theta \\
          =\, & h^2 + y_0^2 -2y_0 \sin \theta~,
\end{equa}
Using the law of reflection; \ie
$- \hat \bt_i \cdot {\hat \bn} = {\hat  \bn} \cdot \hat \bt_e$, we find
\begin{equ}
\frac{D \cos \theta + y_o \sin \theta -1}{h_i} = \frac{D \cos \theta - 1}{h}~,
\end{equ}
from which it follows that
\begin{equ}
y_0 = \frac {2D \sin \theta\ (D \cos \theta - 1)}{(D \cos 2\theta - \cos \theta)}~.
\label{eqn:z0}
\end{equ}
The vanishing of the denominator, resulting in $y_0 \rightarrow \infty$, is equivalent to the incoming ray being normal to the symmetry axis.


\subsection {The $R$-Viewable Surface}
The $R$-viewable surface containing points of radial focus are, as in Section~\ref{sec:Infinite}, computed using the one-dimensional set of rays emanating from $P$ that are in the plane of \fref{fig:FlatConfig}. Consider a ray reflecting from ${\hat  \bn}^{\prime} = \left[ \sin (\theta + \varepsilon) \underline{i}, \ \cos (\theta + \varepsilon)\right]$. The direction of the incoming ray is
\begin{equ}
\hat \bt_i^{\prime} = \frac{1}{h_i^{\prime}}\left[ (\sin (\theta + \varepsilon) - y_0)\ \undi, \ (\cos (\theta + \varepsilon) - D) \right]~,
\label{eqn:new1_inray}
\end{equ}
with
\begin{equ}
h_i^{\prime 2} = 1 + D^2 + y_0^2 -2D \cos (\theta + \varepsilon) - 2 y_0 \sin(\theta + \varepsilon)~, 
\end{equ}
and hence
\begin{equ}
h_i^{\prime} \ \frac{\partial h_i^{\prime}}{\partial \varepsilon} \Bigg|_{\varepsilon=0} =  D \sin \theta - y_0 \cos \theta ~.
\label{eqn:hi_deriv}
\end{equ}
From \eref{eqn:LawofReflection}, the direction of the reflected ray is
\begin{equa}
\hat \bt_e^{\prime} = \frac{1}{h_i^{\prime}} \bigl[ (& -\sin (\theta + \varepsilon) + D \sin 2(\theta + \varepsilon) - y_0 \cos 2(\theta + \varepsilon))\ \undi, \\ & (-\cos (\theta + \varepsilon) + D \cos 2(\theta + \varepsilon) + y_0 \sin 2(\theta + \varepsilon) ) \bigr]~,
\end{equa}
and hence 
\begin{equa}
\frac{\partial \hat \bt_e^{\prime}}{\partial \varepsilon} \Bigg|_{\varepsilon=0} = &\frac{1}{h_i} \left[ (-\cos \theta + 2D\cos 2\theta + 2y_0 \sin 2\theta) \ \undi, \ (\sin \theta - 2D \sin 2\theta + 2y_0 \cos 2\theta) \right] \\
 &- \frac{1}{h_i^3}(D\sin \theta - y_0 \cos \theta) \cdot \hat \bt_e^{\prime} \big|_{\varepsilon=0}~.
 \end{equa}
 
\begin{figure}
\C{\includegraphics[width = 0.60\textwidth]{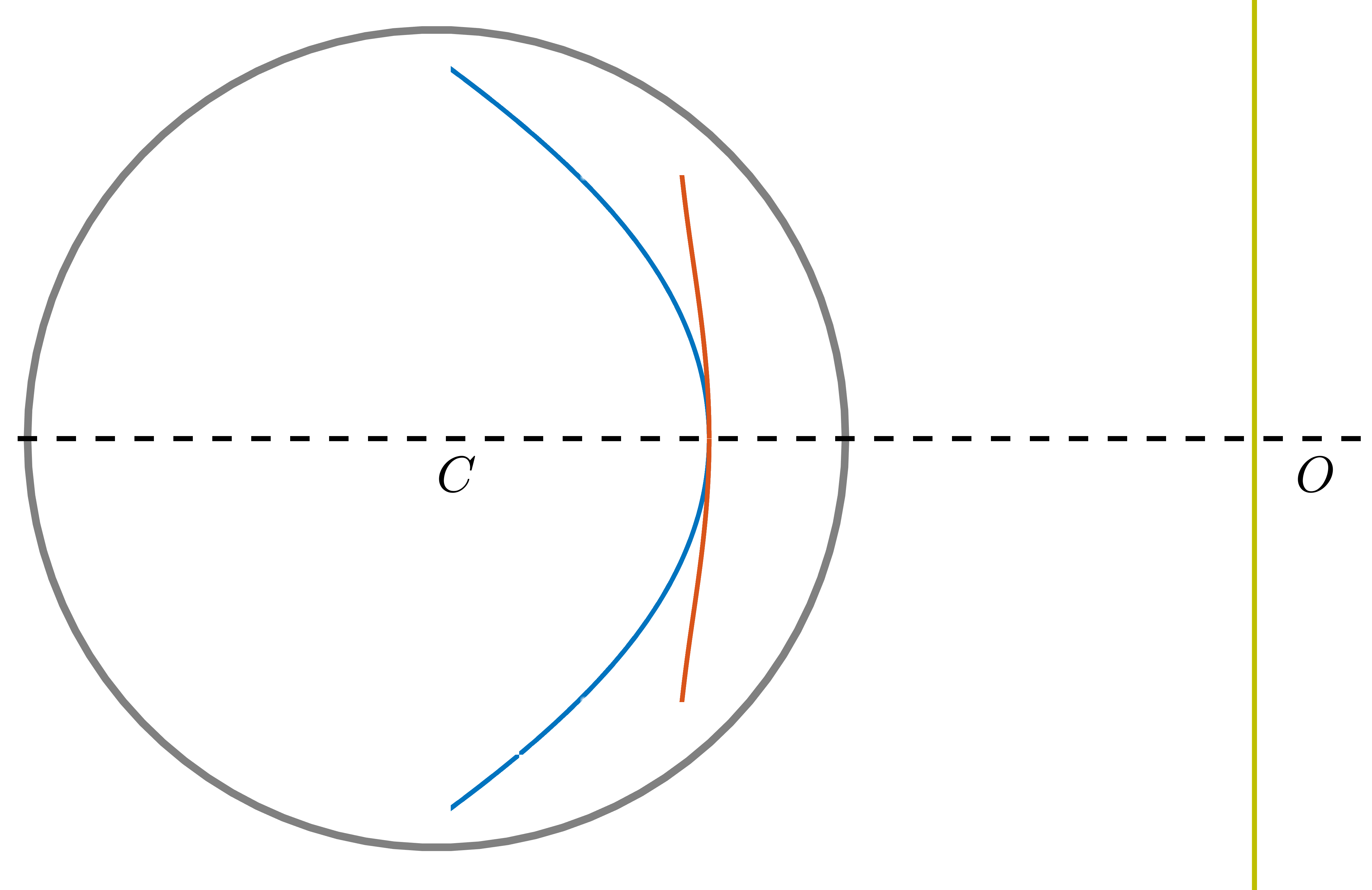}}
\caption {Cross sections of $R$- and $A$-viewable surfaces (red,
  resp.~blue) for a
  planar object placed normal to the axis at  $D = 2R$, and viewed
  through a small hole at its center. }
\label{fig:FlatView} 
\end{figure}

 To compute the viewable surface, note, as before that the reflected ray can be parameterized as $\br_R = {\hat \bn}^{\prime} + s \hat \bt_e^{\prime}$, and that on the viewable surface
 \begin{equ}
 \d\br_R = \frac{\partial {\hat \bn}^{\prime}}{\partial \varepsilon}
 \Bigg|_{\varepsilon=0} \d\varepsilon +  \hat \bt_e^{\prime} \big|_{\varepsilon=0} \d s+ s \frac{\partial \hat \bt_e^{\prime}}{\partial \varepsilon} \Bigg|_{\varepsilon=0} \d\varepsilon = 0~,
 \end{equ}
 from which it follows that the location on the surface is
 \begin{equ}
 s_R = \frac {h_i (D\cos \theta - 1) } {(3D \cos \theta - 1 + 2y_0 \sin \theta - 2D^2 \cos 2\theta - 2 y_0 D \sin  2\theta)}~.
 \end{equ}
The coordinates $y_R$ and $z_R$ on the viewable surface are found to be
\begin{equa}
z_R &= \cos \theta + s_R \frac{(D - \cos \theta)}{h}~;\\
y_R &= \sin \theta - s_R \frac{\sin \theta }{h}~.
\label{eqn:RViewSurface}
\end{equa}
 The cross-section of the $R$-viewable surface for a planar
 object placed a distance $2R$ in front of the witch ball is shown in
 red in \fref{fig:FlatView}. Although it is nearly flat, the surface fails to have a vanishing second derivative for any $D$.

\subsection {The $A$-Viewable Surface} 

To compute the $A$-viewable surface, we need the focus of a rays emanating from $P$ normal to the plane of \fref{fig:FlatConfig}, Consider a ray with a differential orientation $\eta$. The incident direction, point of reflection and other relevant quantities are
\begin{equa}
{\hat \bn}^{\prime} &= \left[ \sin \theta\ {\rm \underline{i e^{i\eta}}},\ \cos \theta \right]~, \nonumber\\
{\hat \bt_i}^{\prime} &= \frac{1}{h_i^{\prime}} \left[ \sin \theta \ {\rm \underline{i e^{i\eta} }} - y_0\ \undi, \ \cos \theta - D \right]~, \nonumber \\
h_i^{\prime 2} &= D^2 + y_0^2 + 1 - 2D \cos \theta - 2 y_0 \sin \theta \cos \eta~, \nonumber \\
\frac{\partial h_i^{\prime}} {\partial \eta} & \big|_{\eta=0} = 0~, \nonumber \\
\hat \bt_e^{\prime}  \big|_{\eta=0} &= \frac{1}{h_i} \left[ (-\sin \theta -y_0 \cos 2\theta + D \sin 2\theta)\ \undi,\ (-\cos \theta + y+0 \sin 2\theta + D \cos 2\theta) \right]~,\\
\frac{\partial \hat \bt_e^{\prime}} {\partial \eta} & \big|_{\eta=0} = \frac{1}{h_i} (-\sin \theta - 2 y_0 \sin^2 \theta - D \sin 2\theta) \left[\underline{1},\ 0 \right]~.
\end{equa}
The reflected ray, parameterized, is $r_A = {\hatt \bn}^{\prime} + s \bt_e^{\prime}$, and the location on the viewable surface corresponds to   
\begin{equ}
s_A = -\frac {h_i}{(2D \cos \theta + 2 y_0 \sin \theta -1)}~,
\end{equ}
and consequently,
\begin{equa}
z_A &= \cos \theta + s_A \frac{(D -\cos \theta)}{h}~,\\
y_A &= \sin \theta - s_A \frac{\sin \theta}{h}~.
\label{eqn:AViewSurface}
\end{equa}
The cross-section of the $A$-viewable surface is shown in blue in \fref{fig:FlatView}.

In the next section we experimentally demonstrate the existence of two distinct images along a viewing direction and validate the expressions \eref{eqn:RViewSurface} and \eref{eqn:AViewSurface}.

\section {Experimental Validation}
\label{sec:Expt}

\begin{figure}
\C{\includegraphics[width = 0.60\textwidth]{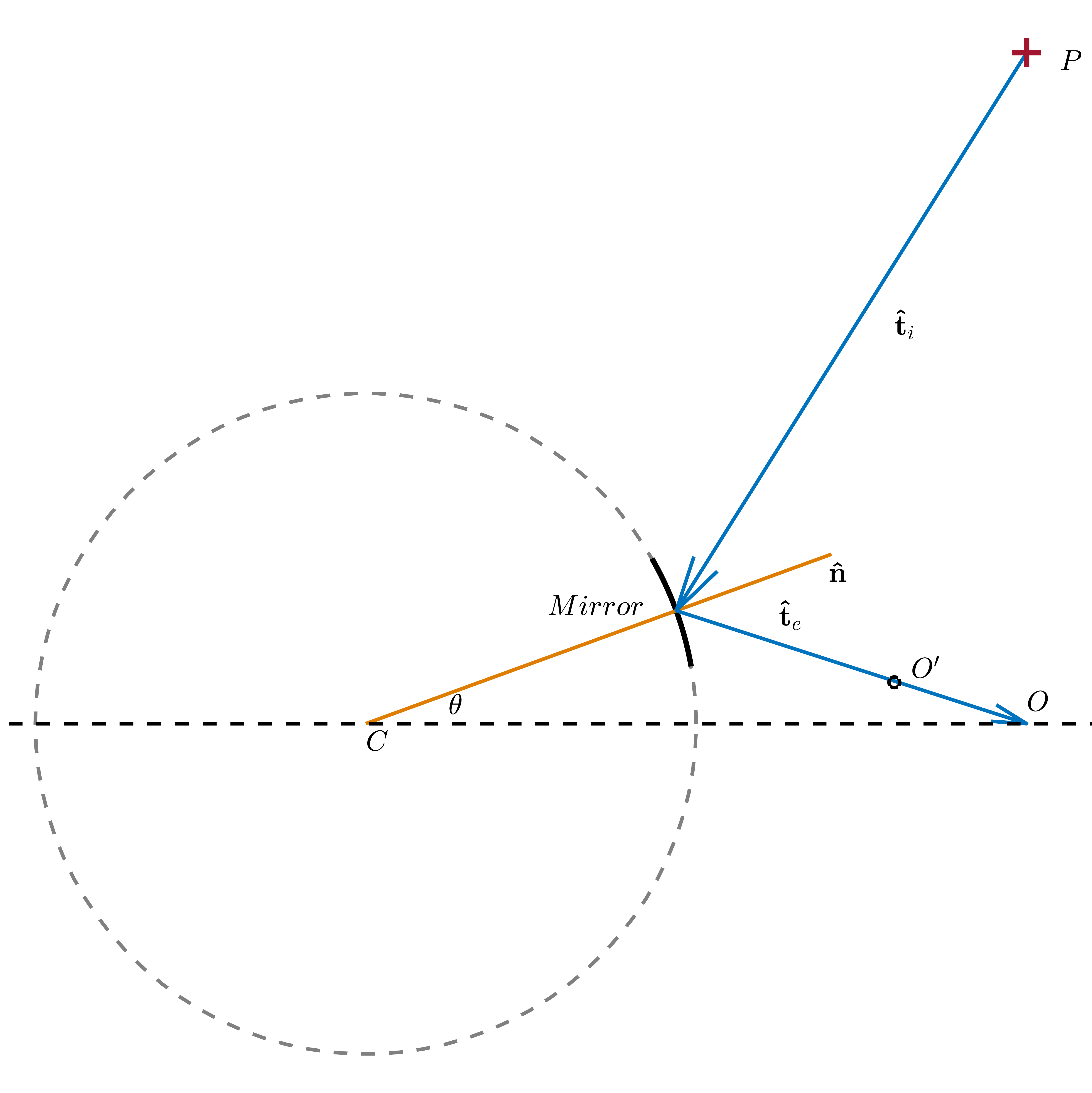}}
\caption {A schematic of the experiment.  The mirror, shown by a solid line is a section of a hypothetical witch ball centered at $C$ is marked by a dashed line. A cross of vertical and horizontal  lines at $P$ is viewed through the camera at $O$. The camera can be moved along $O O^{\prime}$, the direction of the emergent ray.}
\label{fig:Experiment} 
\end{figure}

Typical witch balls are not sufficiently smooth to test the calculations outlined in Section~\ref{sec:ViewingPlane}. 
Experiment validation was conducted using the arrangement shown schematically  in \fref{fig:Experiment}.  The mirror is an optical quality convex spherical mirror with a radius of curvature $R=20.0$ cm, a focal length of $f = -10.0 $ cm, a diameter of $50.0$ mm, and a protected aluminum coating. The center of the sphere is $C$. The object at $P$ is a cross with vertical and horizontal line segments of length $6$ cm and width $2$ mm.  Images are recorded using a Scion camera with a $ccd$ chip of $1200\ {\rm pixels} \times 1600\ {\rm pixels}$  with pixel spacing $4.0 $ micrometers.  A Navitar Zoom 7000 18 --108 mm Macro Lens is used to focus the reflected object onto the detector.  Camera positioning was accomplished using translation stages having micrometer drives with a spatial resolution of $10$ micrometers.  The configuration shown lies in a horizontal plane. 

\begin{figure}
\C{\includegraphics[width = 0.80\textwidth]{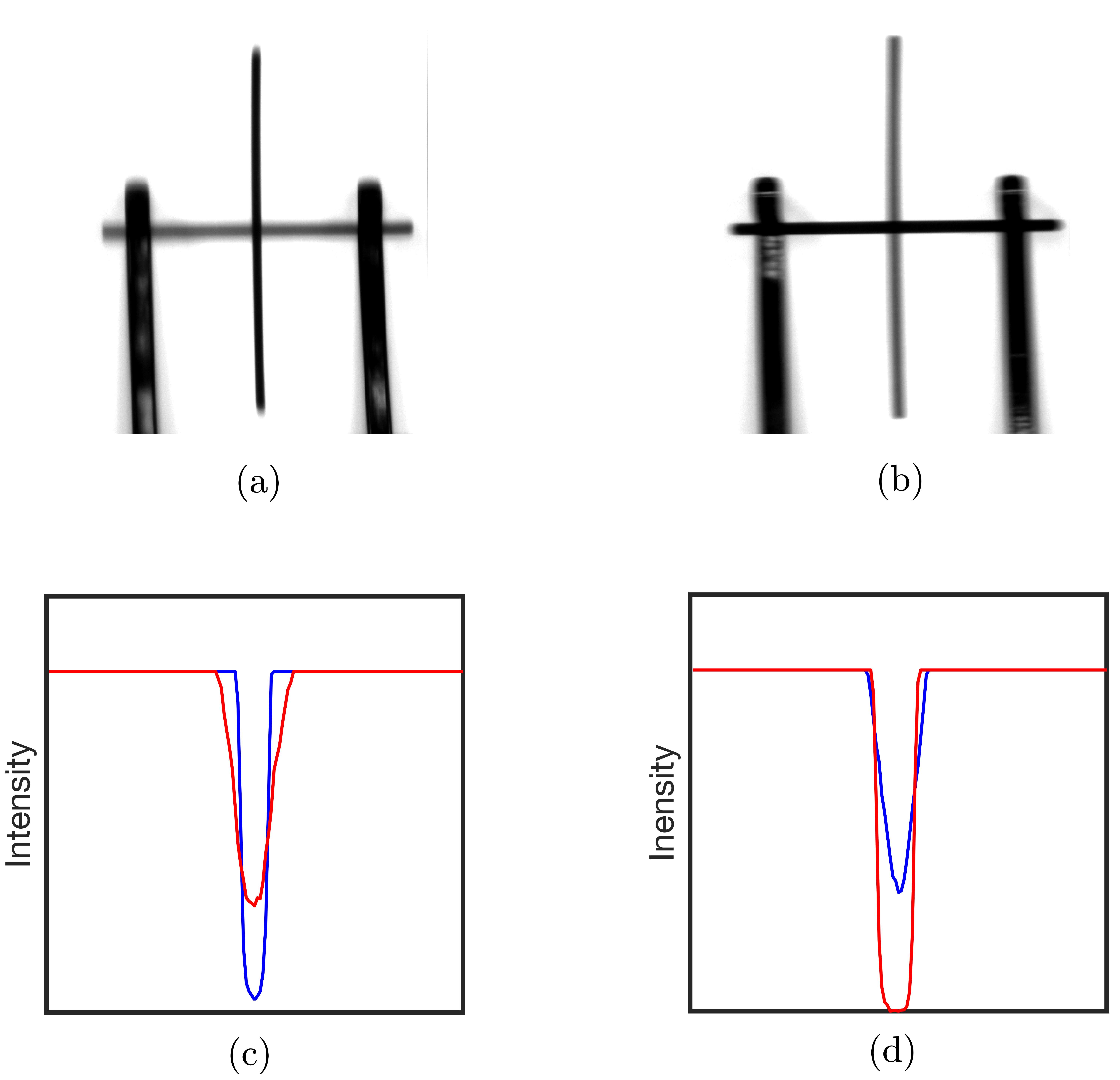}}
\caption {The two images with (a) a sharp focus on the azimuthal (vertical) line, and (b) sharp focus on the radial  (horizontal) line. The variation in intensity for the two images are shown in (c) and (d). The red (resp. blue) line shows the intensity profile for the radial (resp. azimuthal)  lines. Observe that the focussed images have a higher value for the peak intensity gradient. }
\label{fig:TwoFocuses} 
\end{figure}

The mirror and the witch ball from which it is assumed to be cleaved are shown in solid and dashed lines respectively. The mirror-axis is held in place, and each setting is initiated by a pre-specified rotation of the mirror. The center of the sphere,  $C$, is identified using the radius of curvature; the camera is placed on the symmetry axis a distance $D= 65$ mm from $C$. The object $P$ is moved in the plane normal to the symmetry axis until its image is at the focal plane of the mirror. The azimuthal line  (\ie vertical line  in our setup) is set in focus, as confirmed using the intensity profile of the cross section, see \fref{fig:TwoFocuses}(c). The best focus is defined by minimizing the half-width of the intensity curve shown in \fref{fig:TwoFocuses}.(c) Typically, this optimal focus remains unchanged when $O$ is moved by $\sim 1-2$ mm;  these distances give an estimate of the measurement error. Generally, the image of the radial (horizontal) line is not in focus at this point as seen in \fref{fig:TwoFocuses}(c).

Refocussing the camera to locate the second image will change the focal plane in a way that is difficult to estimate. Instead, we search for the second image by moving the camera forward, say by a distance $\delta$, along the direction $-{\hat {\bf t}}_e$ of the exit ray while maintaining the same focus. It is found that the horizontal line of the object focusses when the camera has moved to a point $O^{\prime}$;  $\delta = |O O^{\prime}|$. The use of the micrometer permits the estimation of $\delta$ to within 10 microns, if necessary. Generally, the focus of the azimuthal line is lost at this point as seen from \fref{fig:TwoFocuses}(d). The views at $O$ and $O^{\prime}$, shown in \fref{fig:TwoFocuses}, clearly illustrates the existence of two distinct images. 

\begin{figure}
\C{\includegraphics[width = 0.50\textwidth]{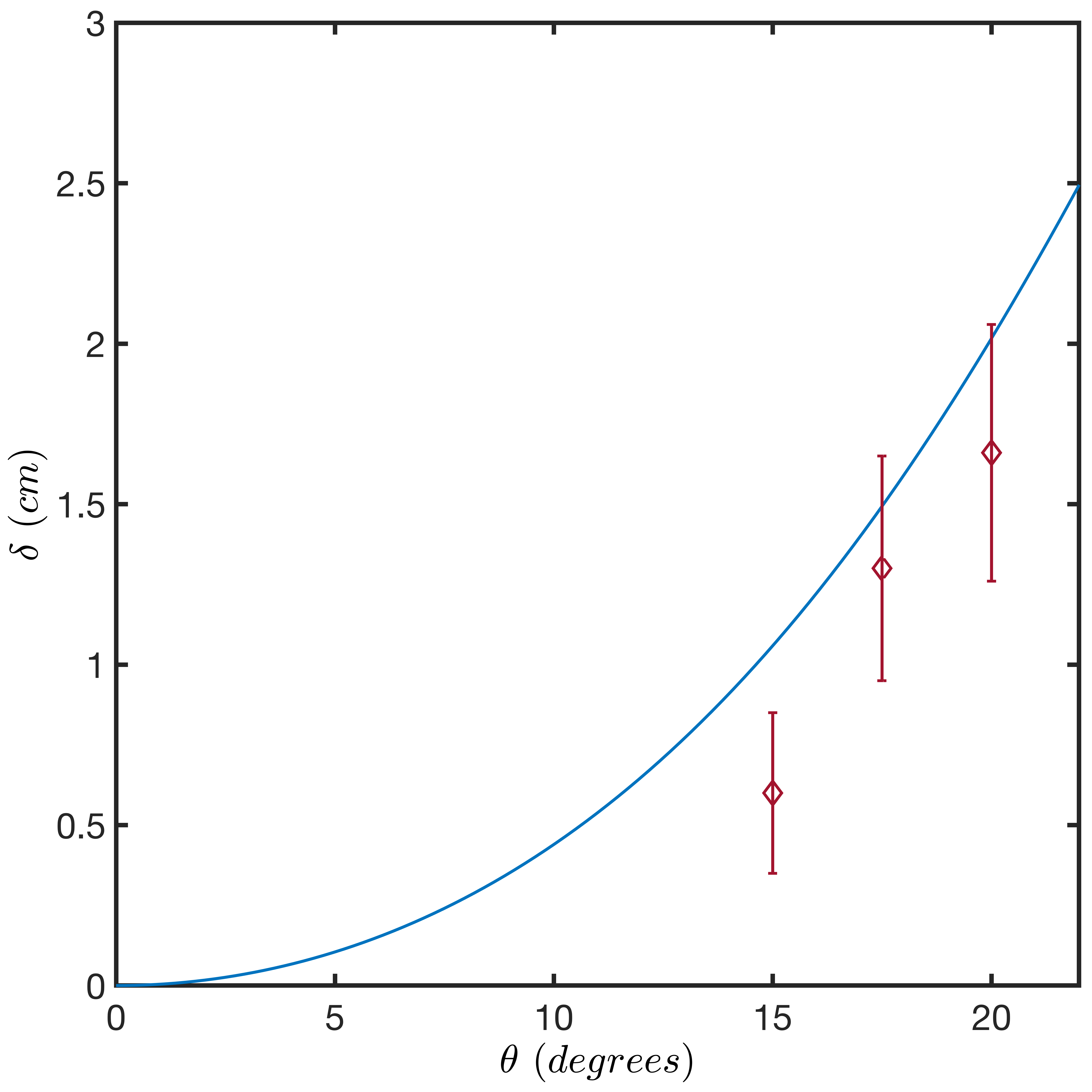}}
\caption {The solid line shows the inclination of the mirror vs. $\delta$, the camera movement to get from the image with vertical focus to that with horizontal focus. The diamonds show the best estimate of $\delta$ and the error bars represent distances over which the with of the intensity remains unchanged.}
\label{fig:Result} 
\end{figure}

Since the camera has moved during the experiment, $\delta$ is not the distance between the $R$ and $A$ viewable surfaces with fixed viewing at $O$. The location of azimuthal focus,  $\br_R(D, \theta)$ is given by ~\eref{eqn:RViewSurface}, while that at radial focus needs to be calculated for viewing at $O^{\prime}$. To do so, we note first that the angle $\psi$ between the exit ray ${\bf t}_e$ and $CO$ is small is our experiments. In addition, $\delta \ll D$ and hence distance $D^{\prime} \equiv |CO^{\prime}| \approx D-\delta$. The angle subtended by $O O^{\prime}$ at $C$ is, by the law of sines, $\delta \theta = \sin^{-1} (\delta \sin \psi / D^{\prime})$. Hence the angle $\theta^{\prime}$ between $CO^{\prime}$ and $\hat {\bf n}$ is $\theta^{\prime} = \theta  + \mathcal{O}(\psi^2, \psi \delta, \delta^2)$; \ie it is unchanged to first order in $\delta$ and $\psi$. The location $\br_A (D^{\prime}, \theta^{\prime})$ of the horizontal focus with the camera at $O^{\prime}$  can be calculated to first order in perturbation in $\delta$ and $\psi$. The value of $\delta$ can then be obtained by solving
\begin{equ}
\delta = | \br_R(D, \theta) - \br_A(D^{\prime}, \theta^{\prime})|.
\end{equ}
The solid line in \fref{fig:Result} shows $\delta$ as a function of $\theta$.  

The experimental values of $\delta$ for $\theta=15^o$, $17.5^o$ and $20^o$ are shown with the error bars originating from our inability to precisely locate the point of best focus.

\section{Discussion}
\label{sec:Discussion}
The role of caustics in image formation was highlighted in Chapter 3 of ~\citep{Feigenbaum2050}. Following passage through a general  (non-symmetric) optical system, a wavefront does not appear to emerge from a single point. Rather, it is a Gaussian surface with partial focus along two principal directions; in ~\citep{Feigenbaum2050}, such wavefronts were considered to emerge from ``non-objects." An observer finds two distinct images of a point source and it was conjectured that our visual system preferentially focusses on the one with vertical astigmatism. In viewing an extended object the two sets of images span two surfaces, referred to as viewable surfaces. In general, these viewable surfaces are not sagittal and tangential surfaces discussed in the context of image formation in axisymmetric optical configurations~\citep{AvendanoAlejoHernandez:2019}. The latter are not computed using caustics, but rather through ray diagrams in two orthogonal planes. Anamorphic images and the corresponding viewable surfaces were calculated in~\citep{Feigenbaum2050}. Imaging of underwater objects was discussed in ~\citep{HorvathBucha:2003,Eckmann:2021}.

The goal of our work was to compute images and viewable surfaces in reflections from a spherical mirror. We computed them for three configurations, the first of which was for an incoming parallel beam of light~\citep{Berry:1972,Berry2015}. The two images and the corresponding viewable surface contained points of radial and azimuthal focus. The surface of radial focus was azimuthally symmetric with a cusp singularity on the symmetry axis. The surface of azimuthal focus degenerated to a one-dimensional subset of the symmetry axis. Both viewable surfaces for the next two configurations were non-degenerate. In one we calculated viewable surfaces for an observer at a distance $D$ viewing infinitely far away objects. We found that the surface of radial focus was planar to fourth order when $D = (2+\sqrt{3}) R$. The final example involved viewing of a planar object through an opening at its intersection with a radius of the sphere normal to the plane. Several predictions were experimentally validated. 

The two viewable surfaces in all configurations we study have the following feature: one surface consisted of points of radial focus and the other points of azimuthal focus. Feigenbaum conjectured that the preferential focus for our visual system is for points of horizontal focus or vertical astigmatism. This leads to an interesting question on viewing a plane in the configuration shown in \fref{fig:FlatConfig} . The viewable surfaces, shown in \fref{fig:FlatView}, contain points of radial (red) and azimuthal (blue) focus. The first viewable surface can easily be captured by a camera. What will a subject observe in viewing the image in the mirror? Will they see the surface of radial focus at the upper and lower locations and the surface of azimuthal focus at the sides? Alternatively, will our ability to reconstruct local scenes using global properties~\citep{GreeneOliva:2009} allow a subject to focus on one of the two images?  We expect to conduct a study to address this issue. 

JPE is partially supported by the Fonds National Suisse Swissmap. The work reported here was inspired by studies of Mitchell J. Feigenbaum and insightful discussions with him. We dedicate this work to his memory.

%
%

%
%

\bibliographystyle{new4}
\bibliography{OpticsRefs}

\begin{thebibliography}{12}
\providecommand{\natexlab}[1]{#1}
\providecommand{\url}[1]{\texttt{#1}}
\providecommand{\urlprefix}{URL }

\bibitem[{Avenda\~no Alejo(2013)}]{AvendanoAlejo:2013}
Avenda\~no Alejo, M. (2013).
\newblock Caustics in a meridional plane produced by plano-convex aspheric
  lenses.
\newblock \emph{Journal of the Optical Society of America A}, 30.

\bibitem[{Avenda\~no Alejo et~al.(2019)Avenda\~no Alejo, Roman-Hernandez,
  Castillo-Santiago, DelOlmo-Marquez, and
  Caste\~neda}]{AvendanoAlejoHernandez:2019}
Avenda\~no Alejo, M., Roman-Hernandez, E., Castillo-Santiago, G.,
  DelOlmo-Marquez, J., and Caste\~neda, L. (2019).
\newblock Sagittal and tangential foci produced by tilted plane wavefronts
  through simple lenses.
\newblock \emph{Applied Optics}, 58.

\bibitem[{Bartlett et~al.(1984)Bartlett, Lucero, and
  Johnson}]{BartlettLucero:1984}
Bartlett, A.~A., Lucero, R., and Johnson, G.~O. (1984).
\newblock Note on a common virtual image.
\newblock \emph{American Journal of Physics}, 52:640--643.

\bibitem[{Berry(1972)}]{Berry:1972}
Berry, M.~V. (1972).
\newblock Reflections on a christmas-tree bauble.
\newblock \emph{Physics Education}, 7:1--6.

\bibitem[{Berry(2015)}]{Berry2015}
Berry, M.~V. (2015).
\newblock The squint moon and witch ball.
\newblock \emph{New Journal of Physics}, 17.

\bibitem[{Eckmann(2021)}]{Eckmann:2021}
Eckmann, J.-P. (2021).
\newblock Broken pencils and moving rulers: After an unpublished book by
  {M}itchell {F}eigenbaum.
\newblock \emph{American Journal of Physics}, 89:955--962.

\bibitem[{Feigenbaum(unpublished)}]{Feigenbaum2050}
Feigenbaum, M.~J. (unpublished).
\newblock \emph{Reflections on a Tube}.
\newblock Edited by Jean-Pierre Eckmann.

\bibitem[{Greene and Oliva(2009)}]{GreeneOliva:2009}
Greene, M.~A. and Oliva, A. (2009).
\newblock Recognition of natural scenes from global properties: Seeing the
  forest without representing the trees.
\newblock \emph{Cognitive Psychology}, 58:137--176.

\bibitem[{Horv\'ath et~al.(2003)Horv\'ath, Buchta, and
  Varj\'u}]{HorvathBucha:2003}
Horv\'ath, G., Buchta, K., and Varj\'u, D. (2003).
\newblock Looking into the water with oblique head tilting: revision of the
  aerial binocular imaging of underwater objects.
\newblock \emph{Journal of the Optical Society of America}, 20:1120--1131.

\bibitem[{Kinsler(1945)}]{Kinsler:1945}
Kinsler, L.~E. (1945).
\newblock Imaging of underwater objects.
\newblock \emph{American Journal of Physics}, 13:255--257.

\bibitem[{Nussenzveig(1977)}]{Nussenzveig1977}
Nussenzveig, H.~M. (1977).
\newblock The theory of the rainbow.
\newblock \emph{Scientific American}, April:116--127.

\bibitem[{Nye(1999)}]{Nye:1999}
Nye, J.~F. (1999).
\newblock \emph{Natural Focusing and Fine Structure of Light: Caustics and Wave
  Dislocations}.
\newblock Institute of Physics Publishing.

\end{thebibliography}



\end{document}